\documentclass[twocolumn,showpacs,amssymb,aps,nofootinbib,floatfix]{revtex4}
\usepackage{epsfig}
\newcommand{\be}{\begin{eqnarray}}
\newcommand{\ee}{\end{eqnarray}}

\begin{document} \hbadness=10000
\topmargin -0.8cm\oddsidemargin = -0.7cm\evensidemargin = -0.7cm
\title{Hadron Resonances and Phase Threshold in Heavy Ion Collisions}
\author{Giorgio Torrieri}
\affiliation{
Department of Physics, McGill University, Montreal, QC H3A-2T8,
Canada \footnote{Present affiliation: Theoretical Physics,
  J.W. Goethe Universitat, Frankfurt A.M., Germany}}
\author{Johann Rafelski}
\affiliation{
Department of Physics, University of Arizona, Tucson AZ 85721, USA}
\date{August 25, 2006}  

\begin{abstract}
We show that a measurement of the reaction
energy ($\sqrt{s}$) dependence of relative hadron resonance yields in heavy ion collisions can be 
used to study the phase structure of the dense strongly interacting matter created in these collisions, and investigate the 
origin of the trends observed in the excitation functions of certain soft hadronic observables.
We show that presence of chemical nonequilibrium in light quark abundance imparts a characteristic signature 
on the energy dependence of resonance yields, that differs considerably from what is expected in the equilibrium picture. 
\end{abstract}
\maketitle
\section{Introduction}  
The exploration of the  properties 
of strongly interacting, dense  quark-gluon 
matter, and specifically, of  the equation of state, transport coefficients,
degree of equilibration, and phase structure, and the dependence of these on the 
energy and system size is one of the main objectives of heavy ion research program.
A natural approach to these challenges is the study of soft 
particle multiplicities produced in these reactions. This  provides information
about the system properties when these particles are created (chemical freeze-out conditions),
as well as about bulk matter properties (e.g. entropy) 
which can probe deep into the birth history of the 
fireball.

Statistical mechanics techniques have in this context  
a long and illustrious history~\cite{Fer50,Pom51,Lan53,Hag65}.      
The systematic and quantitative comparison of data to the statistical hadronization  (SH)
model is, however,   a comparatively recent 
field~\cite{jansbook,bdm,barannikova,equil_energy,gammaq_energy,gammaq_size,becattini}.
 A consensus has developed that the SH model can indeed 
fit most, if not all particle yields measured at experiments conducted at
a wide range of energies. Measurements conducted at the GSI Schwerionen Synchrotron
(SIS), BNL's Alternating Gradient Synchrotron (AGS),CERN's Super Proton
Synchrotron (SPS),and BNL's Relativistic
Heavy Ion Collider (RHIC) have successfully been analyzed using SH ansatze.

When this consensus is considered more carefully, we see that in technical
 detail the applied SH models differ regarding
the chemical equilibration condition that is presumed.  
 As a result, it has not as yet been possible to agree statistical
 physics, if any, is responsible for the striking trends observed in the energy
 dependence of some observed hadronic yields.  

In this paper we will indicate that further progress can be made with help of hadron resonances.
Hadron resonances, such as e.g. $\Delta^{++}$ excitation of the proton $p$  differ typically from the 
``stable'' particle by internal structure, rather than chemical quark content. 
Hence within the SH approach their production is 
mostly controlled by the ``temperature'' $T$ at which they are created. 
\section{Equilibrium and non-equilibrium freeze-out condition}  
In the SH model there are two types of chemical equilibrium~\cite{koch1986}.:
all models assume relative chemical equilibrium, but some 
also assume absolute chemical equilibrium which implies the presence of   just the right 
abundances of valance  up, down, and even strange quark pairs.
There are  qualitative differences in the results obtained in the description of 
hadron production with or without using  the hypothesis of absolute chemical
equilibrium:  if the system of produced hadrons is considered to be in absolute 
chemical equilibrium, then   at highest heavy ion reaction energy  
one obtains chemical freeze-out temperature   $T \sim 160-170$ MeV. 
Values  as low as $T \sim 50$ MeV are reported at lowest reaction energies available.

\begin{figure*} 
\epsfig{width=19cm,clip=,figure=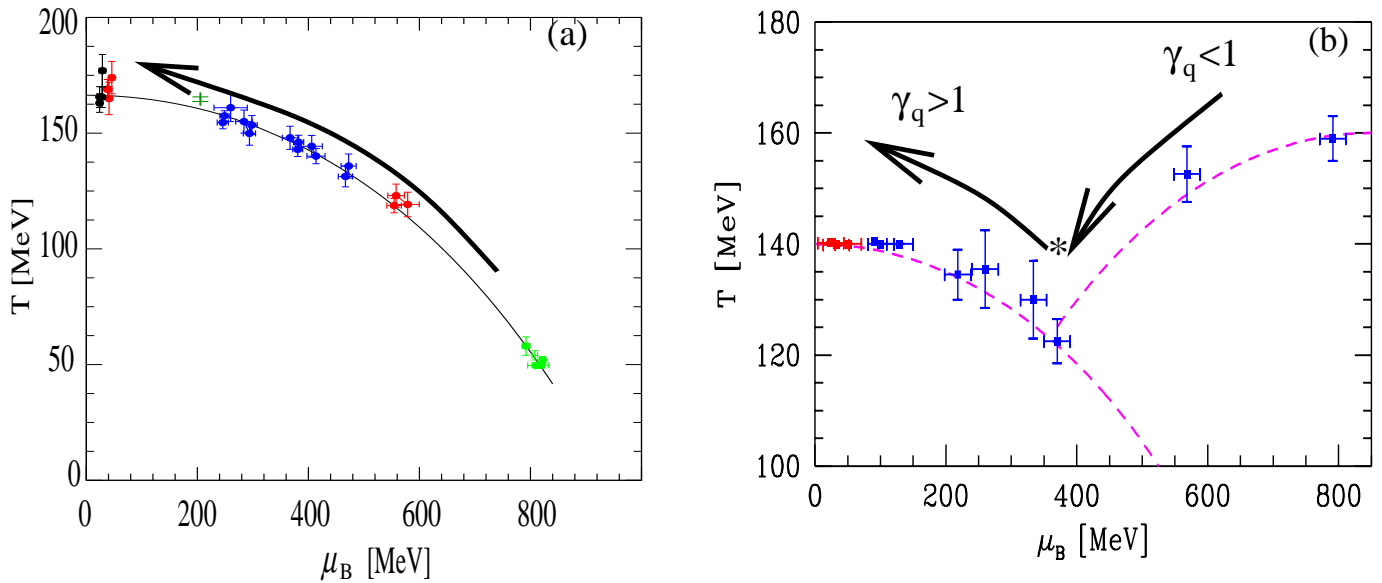}
\vskip -0.5cm
\caption{(Color online)\label{phaseall} 
Dependence of freeze-out temperature  $T$ and baryo-chemical potential $\mu_{\rm B}$
on reaction energy in the Equilibrium 
(panel (a), \cite{equil_energy}) and non-equilibrium (panel (b),\cite{gammaq_energy}) 
freeze-out models.  The direction of the arrow corresponds to increasing
$\sqrt{s}$.  The equilibrium dependence of $T$ and $\mu_B$ in the
 panel (a) is not significantly altered by the introduction of the fitted phase
space occupancy $\gamma_s$ and/or the implementation of the Canonical
ensemble for strangeness.
 The ``star'' in panel (b) corresponds to the point where
the transition to the supercooled  regime occurs and the phase space changes
from chemically under-saturated ($\gamma_q<1$) to chemically over-saturated ($\gamma_q>1$).
This point also corresponds to the energy of the ``kink'' and the tip of the ``horn''}
\end{figure*}

The energy dependence of the freeze-out temperature than follows the
trend indicated in panel (a) of figure \ref{phaseall}: as the collision energy increases, the freeze-out temperature 
increases and the baryonic density (here baryonic chemical potential $\mu_{\rm B}$)
decreases~\cite{equil_energy}.  An  increase of freeze-out temperature with $\sqrt{s}$ is 
expected on general grounds, since  with increasing reaction energy a 
greater fraction of the energy is carried by mesons created in the collision, 
rather than pre-existing baryons~\cite{Hagedorn:1980kb}. 


Further refinements  in the approach  described above
are often implemented and could be of relevance:
\begin{itemize}
\item
Allowance for strangeness chemical nonequilibrium
is necessary  to obtain a good description of strange
particle yields \cite{jangammas,jangammas2,becattini2} at low $\sqrt{s}$.
This is accomplished by introducing
strangeness phase space occupancy $\gamma_s$;\\
\item
At smaller reaction energies and for smaller reaction
system sizes, it is likely that the
fireball is  well away from the thermodynamic limit.
In this case canonical treatment of strangeness is
applied \cite{canon1,canon2}
\end{itemize}
These effects do not materially alter the behaviour of
temperature and chemical potential shown in the panel (a) of
Fig. \ref{phaseall}.

 What is most striking 
in these results is that there is no
sign of any structure when the reaction energy varies. However, 
there are non-continuous  features in the energy dependence of hadronic observables, 
such as the ``kink'' in the multiplicity per number of participants and the 
``horn''~\cite{horn,horn_theory,gammaq_energy,hornthermal} in certain particle yield ratios.   
An effort was made to interpret this in terms of a shift from baryon to meson dominance~\cite{hornthermal}
of the hadron yields. However no matter how hard one
tries,  in the chemical equilibrium model even the simple observable 
like $K^+/\pi^+$ remains a smooth function of reaction energy, in contrast to the
experimental results.   
Introduction of $\gamma_s$ and deviations from
the thermodynamic limit, while they help in bringing some of the model
predictions closer to the data, has so far not managed to reproduce
the sharpness of features such as the kink and the horn.

Non-monotonic behaviour of particle yield ratios could  indicate
a novel reaction mechanism, e.g. onset of the deconfinement 
phase~\cite{horn_theory}.  In such a situation, the smoothness of the chemical 
freeze-out temperature dependence on energy would be surprising, since it would
imply that  at all energies, from about 1 $A$ GeV at SIS, to the highest RHIC values,
there is no change in either  the fireball evolution dynamics, nor any other  imprint from the 
deconfined phase on the freeze-out condition, which, however is visible in 
the strangeness and entropy yield that $K^+$ and, respectively, $\pi^+$ represent.  

Furthermore, we note that the fireball of hadronic matter formed is  
a relatively small system, expanding rapidly, with its content  undergoing a 
phase transformation, or even phase transition. In this complex and rapidly 
evolving circumstance, one could imagine that 
the absolute chemical equilibrium,  not always, or even ever, holds.
In particular, if the expanding system undergoes a fast conversion
from a Quark Gluon Plasma (QGP) to hadrons,
chemical non-equilibrium~\cite{jansbook}, and super-cooling~\cite{sudden,csorgo} go hand in 
hand,  due  to entropy and flavor conservation requirements. 

One can look at this situation again removing the hypothesis of absolute chemical equilibrium among hadrons 
produced. The   systematic behaviour of 
$T$ with energy in this case is quite different~\cite{gammaq_energy},
as  is shown  in panel (b) of
figure \ref{phaseall}.   
The two higher $T$ values  at right are for 20 (lowest SPS) and  (most to right) 11.6 $A$ GeV (highest AGS)
reactions. In these two cases the source of particles is  a hot chemically under-saturated  ($T \sim 170$ MeV ) fireball.
Such a system could be a conventional hadron gas fireball that 
had not the time to chemically equilibrate. Other options were   
 considered in Refs.~\cite{agspaper,gammaq_energy}, such as a phase of constituent massive quarks.

Following the thick arrow in panel (b) of
figure \ref{phaseall} we note that  somewhat smaller temperatures are found with further increasing 
heavy ion reaction energies. 
Here it is possible \cite{gammaqphys,sudden}to match the entropy of the emerging 
hadrons with that of a system of nearly massless partons when one considers supercooling
to $T\sim 140$ MeV,  while both light and strange quark phase space in the hadron stage 
acquire significant over-saturation with the phase space occupancy 
$\gamma_{q=u,d}>1$ and at higher energy also $\gamma_s>1$.  A drastic change 
 in the non-equilibrium condition occurs near 30 $A$ GeV,
corresponding to the dip point on right in panel (b) of the figure \ref{phaseall} (marked by an asterisk). 
At heavy ion reaction energy below (i.e. to right in panel (b) of
figure \ref{phaseall}) of this point, hadrons have not reached 
chemical equilibrium, while   at this point, as well as, at heavy ion reaction energy above 
(i.e. at and to left in panel (b) of figure \ref{phaseall}),   hadrons  emerge from a much denser and chemically more saturated system,
as would be expected were QGP formed at and above 30 $A$ GeV.   
This is also  the heavy ion reaction energy corresponding to the ``kink'', which tracks the QGP's entropy density 
(higher w.r.t. a hadron gas), and the peak of the ``horn'' \cite{horn}, 
which tracks the strangeness over entropy ratio (also higher w.r.t. a hadron gas). 

Concluding this discussion, comparing panel (a) with (b)  we see in  figure  \ref{phaseall} a quite
different behaviour. On panel (a), for the chemical equilibrium model, with increase of the 
collision energy following the black arrow, we see  monotonic increase of 
the chemical freeze-out temperature, with no hint
of new physics in a wide range of heavy ion collision energies spanning the range of SIS, AGS, SPS and RHIC. 
On panel (b), we see that when relative and absolute chemical
equilibrium is considered
~\cite{koch1986},
with  yields of individual hadrons satisfying the relative, but not the  absolute chemical equilibrium,
the experimental particle yield data is best described with a temperature profile as function of reaction
energy which is not monotonic. There is   a minimum value of $T$, at the point when the rapid change of the 
chemical composition of produced hadrons is occurring. This is clearly suggestive of a change in the 
reaction mechanism. 
 
 The main reason for the wider acceptance of the equilibrium 
approach $\gamma_i=1$  is its greater simplicity,
there are fewer parameters. Moreover, considering the quality of 
the data the non-equilibrium parameter $\gamma_q$ is not 
necessary to pull the statistical significance above  it's 
generally accepted  minimal value of 5 $\%$.  On the other hand,
the   parameters $\gamma_q$ and $\gamma_s$ were introduced on 
{\em  physical} grounds \cite{koch1986,gammaqphys,sudden}, thus 
these are not   arbitrary fit parameters.   Moreover,  these
parameters, when used in a statistical hadronization  fit, 
converge to theoretically motivated values. They also  
help to explain the trends observed in the energy dependence 
of hadronic observables. Finally, $\gamma_q>1 $   in AA reactions 
describes the enhancement of the  baryon to meson ratio yield 
at RHIC, as compared to elementary interactions, 
which dynamically arises in the  recombination hadronization 
at fixed hadronization temperature.  
\section{Resonance ratios as chemical freeze-out temperature probes}  
In this paper, we propose the energy dependence of the resonance yields
as a possible experimental observable, capable to discriminate  the 
two scenarios, chemical equilibrium and non-equilibrium, and thus to 
establish the need to use   $\gamma_q$ in statistical hadronization
analysis of experimental data. 

 Many strong interaction resonances, a set we denote by the collective symbol $R^*$ (such as 
$K^{*0}(892), $ $ \Delta(1232), $ $\Sigma^* (1385), $ $\Lambda^* (1520), $ $\Xi^*$ (1530) \cite{pdg}) 
carry the same valance quark content  as their ground-state 
counter-parts $R$ (corresponding: $K,$ $N,$ $\Sigma,$ $\Lambda,$ $\Xi$). $R^*$
 typically decay by emission of a pion,  $ R^*\rightarrow R+\pi$.   
Considering the  particle yield ratio $R^*/R$ in the Boltzmann approximation (appropriate for the particles considered), 
we see that all chemical conditions and parameters (equilibrium and non-equilibrium) cancel out, and the 
ratio of yields between the   resonance and it's
ground state is a function of the masses, and the freeze-out temperature, with 
second order effects coming from the cascading decays of other, more massive 
resonances~\cite{jansbook,share}:
\begin{equation}\label{relY}
\frac{N_{R^*}}{N_R} \simeq \frac{g_{R^*} W \left(\frac{ m_{R^*}}{T}\right)
         +\sum_{j \rightarrow R^*}  b_{jR^*}\,g_{j} W \left(\frac{ m_{j}}{T}\right)}
              { g_{R} W \left(\frac{ m_{R}}{T}\right)
        + \sum_{k\rightarrow R}  b_{kR}\, g_{k}W \left(\frac{ m_{k}}{T}\right)}
\label{ratio}
\end{equation}
where $W(x)=x^2K_2(x)$ is the (relativistic) reduced one particle  phase space, $K_2(x)$
being a Bessel function, 
$g$ is the quantum degeneracy, and $b_{jR}$ is the branching ratio of resonance $j$ decaying into $R$.

When we study the results arising from Eq.\,(\ref{relY}),
we consider only strong decay contributions,   weak  decay feed-down, 
such as $\Lambda \rightarrow p$, $\Sigma \rightarrow p$, $\Xi \rightarrow \Lambda$,
and $\Omega \rightarrow \Xi$  has to be  eliminated from the data sample. 
Given that existing  SPS \cite{na49} and RHIC \cite{rhiclow} experiments,  
as well as the planned Compressed Baryonic Matter (CBM) experiment \cite{cbm} have both a tracking 
resolution permitting precise primary vertex cuts (weak decay tracks originate from points
well away from the primary vertex), as well as a momentum resolution capable
of identifying resonances \cite{na49res,fachini,salur}, this requirement is  realistic.

Because of the radically different energy dependence of freeze-out temperature 
in the scenarios of \cite{equil_energy} and \cite{gammaq_energy}, seen in figure  \ref{phaseall},
the prediction for the resonance ratios Eq. (\ref{relY})  vary greatly between these two scenarios.  
In the equilibrium scenario the temperature goes {\em up} with heavy ion reaction energy, and thus 
 the resonance abundance should go smoothly up for all resonances.
On the other hand, the nonequilibrium scenario, with a low temperature arising only in some 
limited reaction energy domain, will lead to   resonance abundance which
should  have a clear dip  at that point, but otherwise remain relatively large. 

\begin{figure*} 
\epsfig{width=8.1cm,clip=,figure=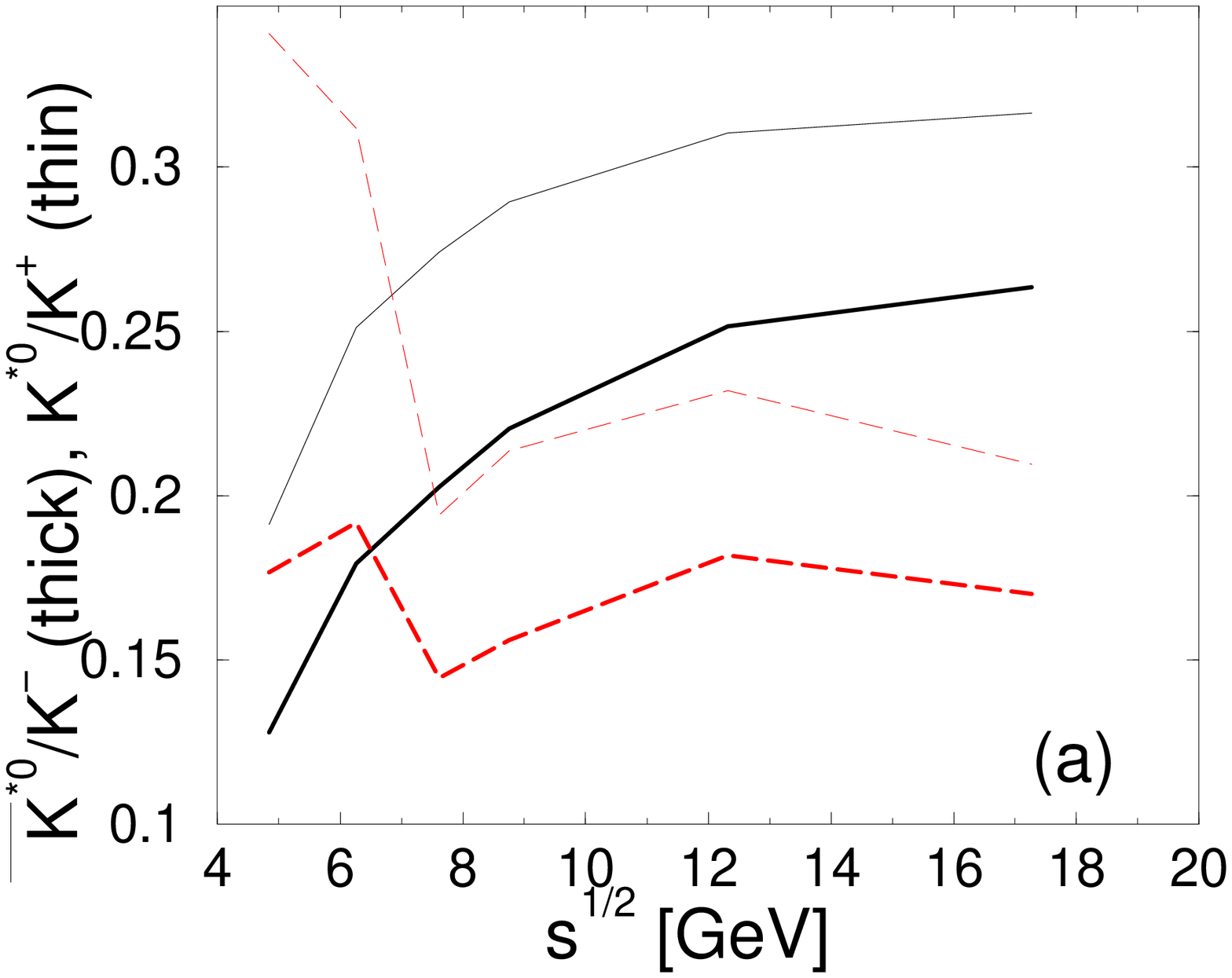}
\epsfig{width=8.1cm,clip=,figure=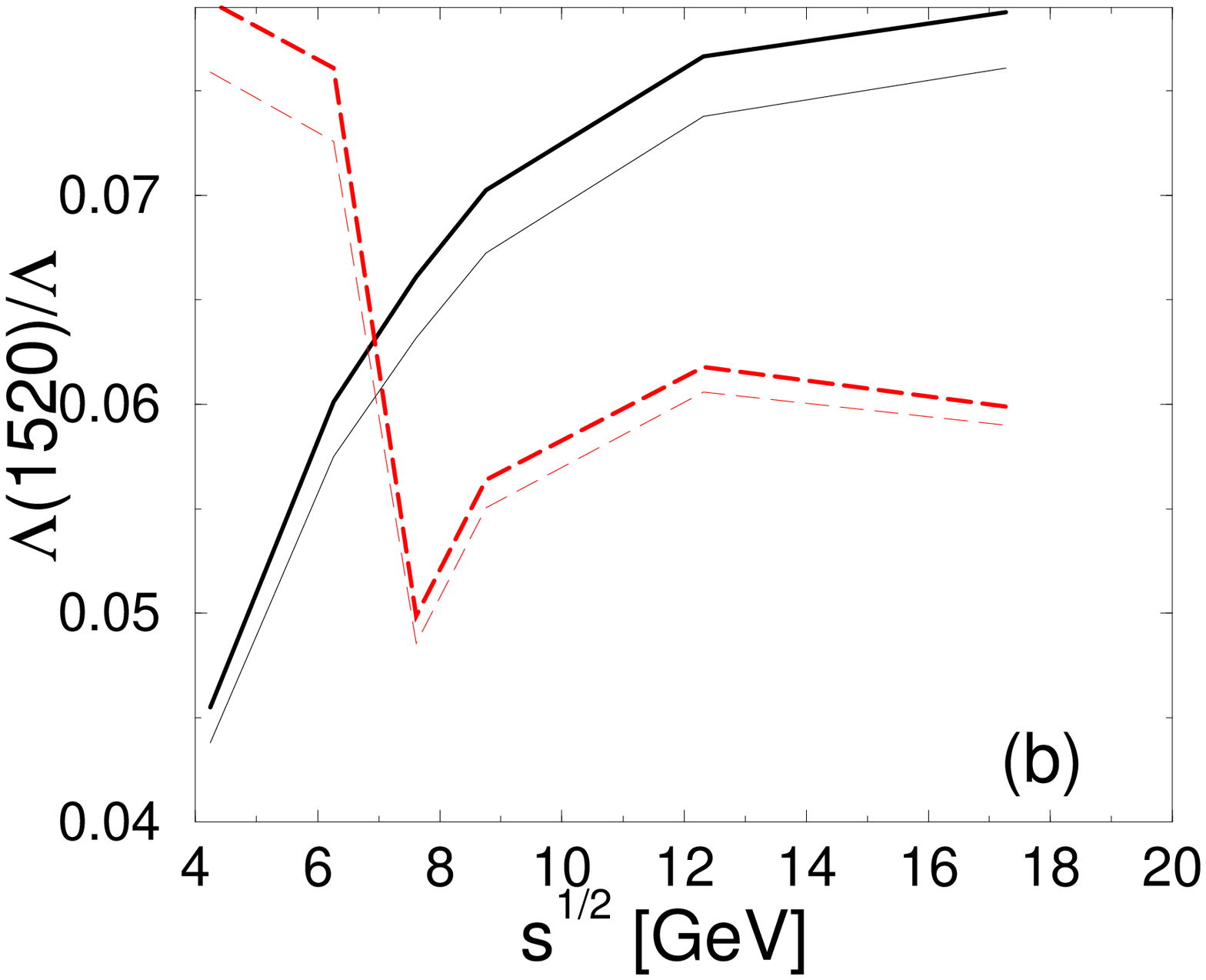}
\epsfig{width=8.1cm,clip=,figure=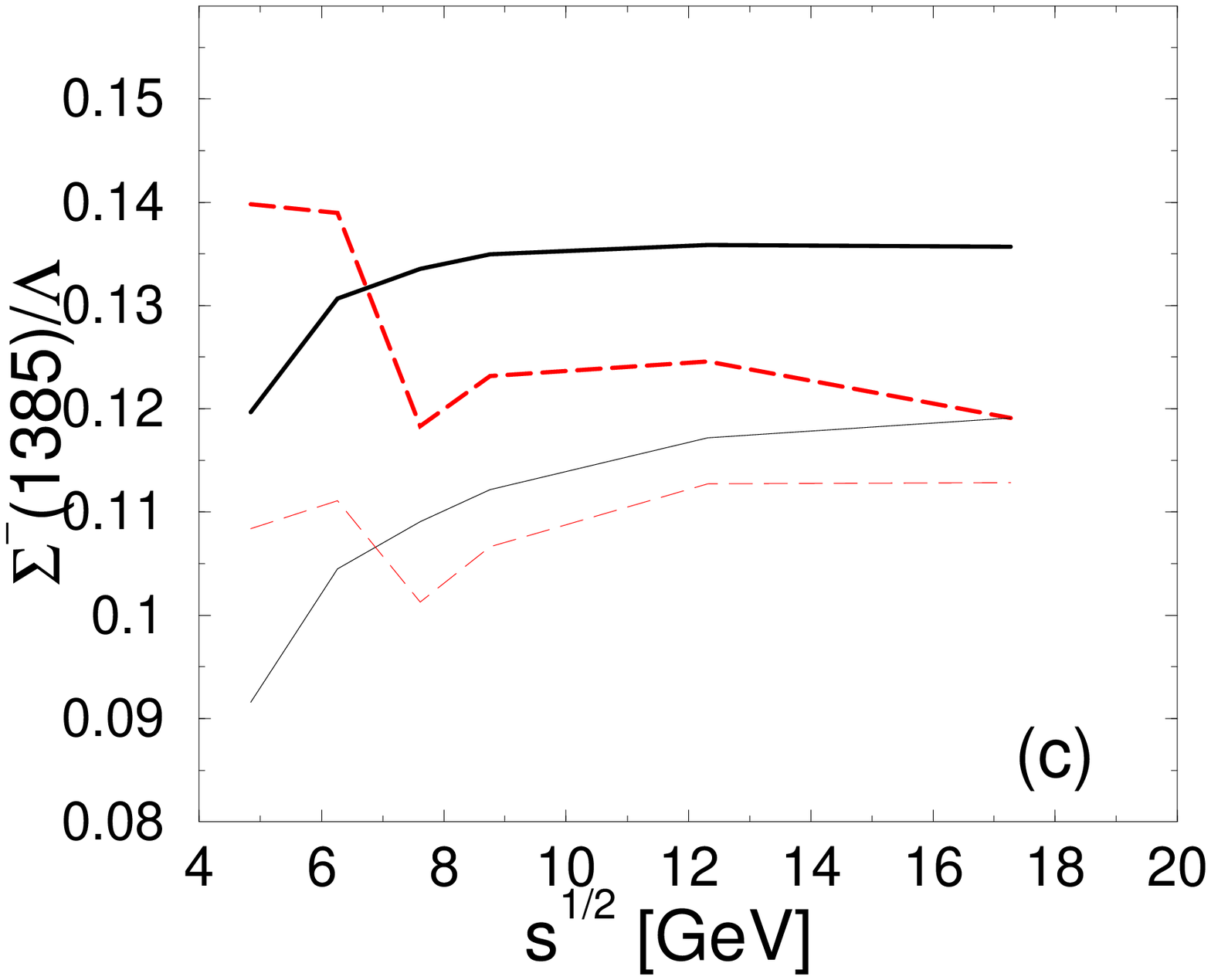}
\epsfig{width=8.1cm,clip=,figure=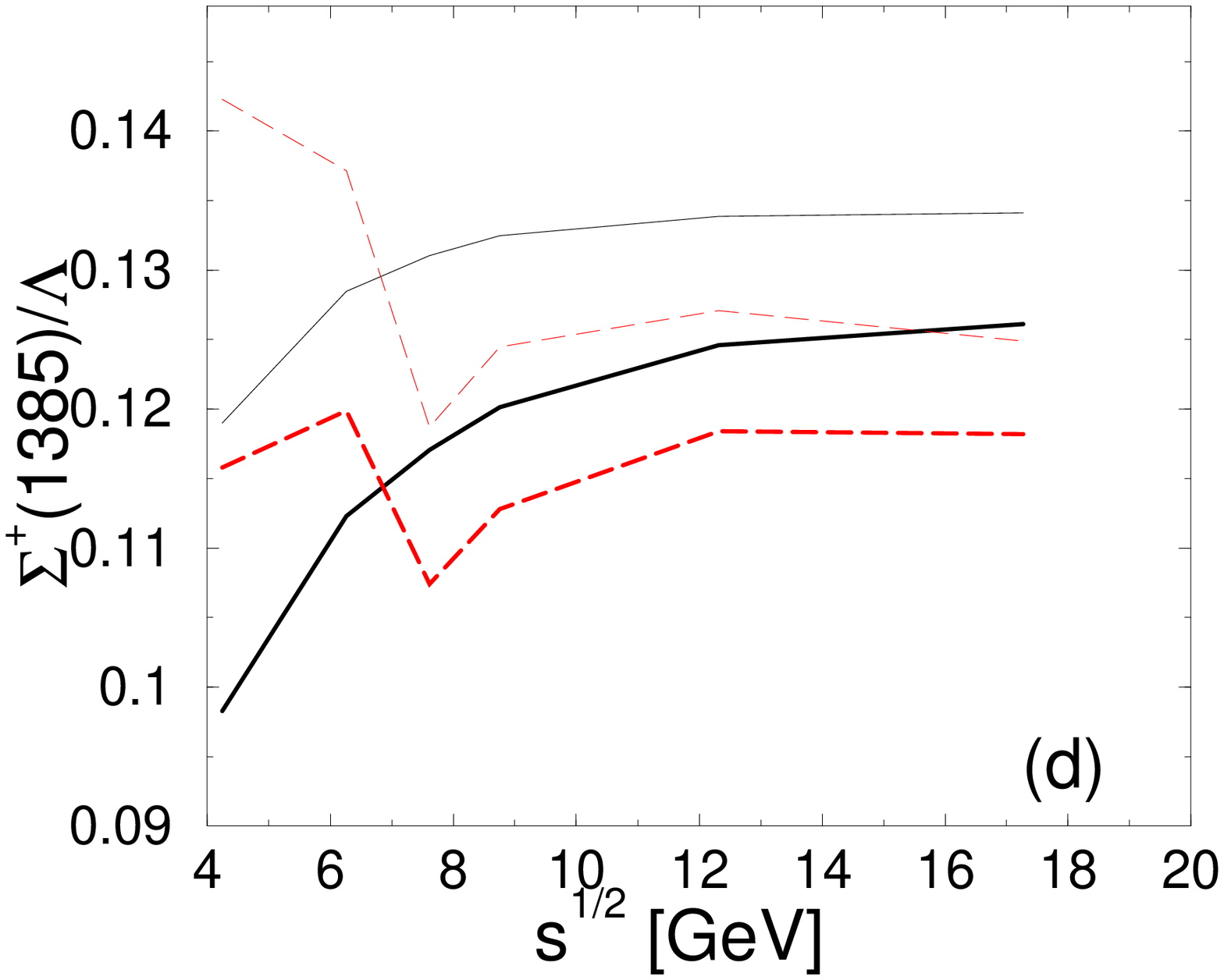}
\epsfig{width=8.1cm,clip=,figure=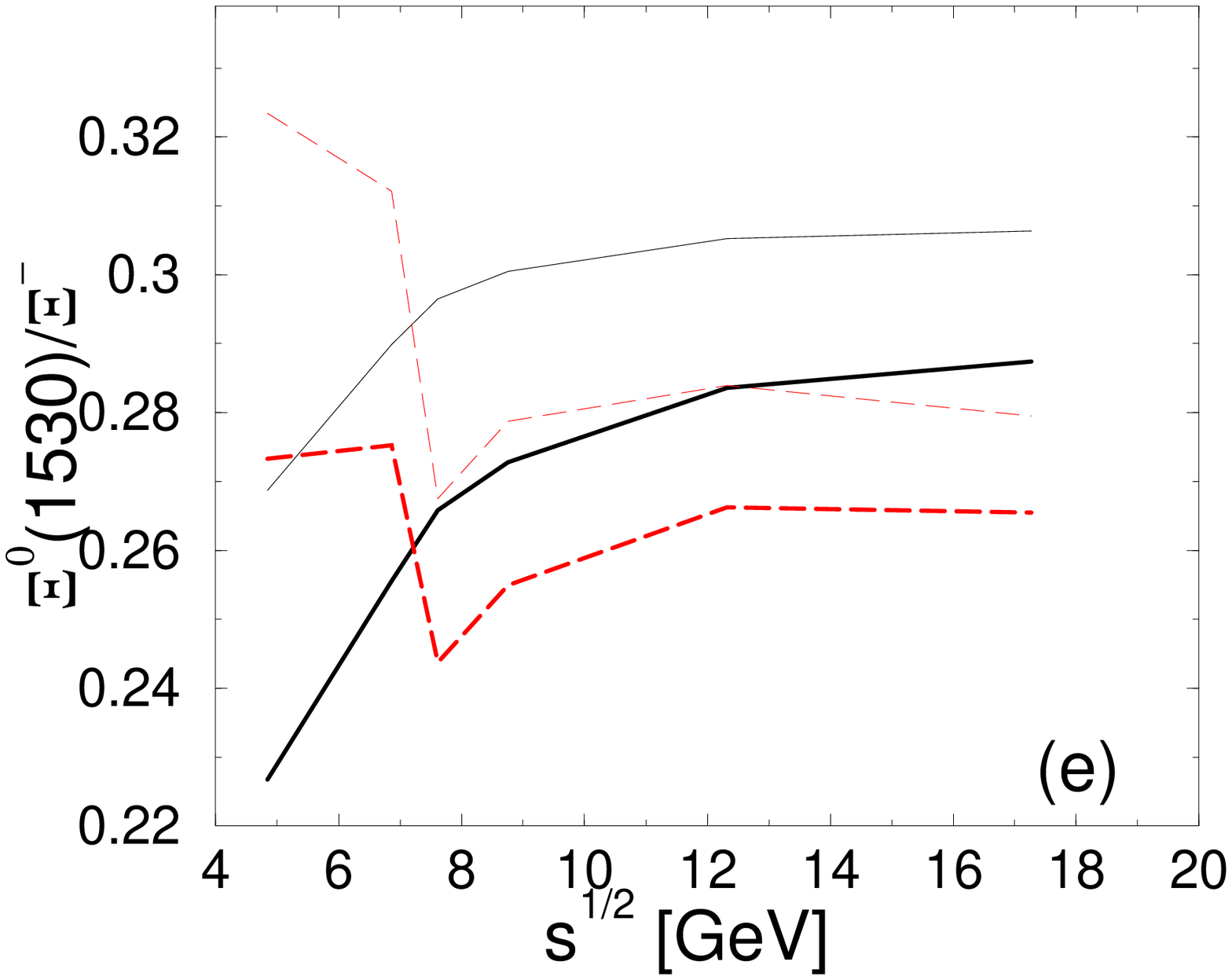}
\epsfig{width=8.1cm,clip=,figure=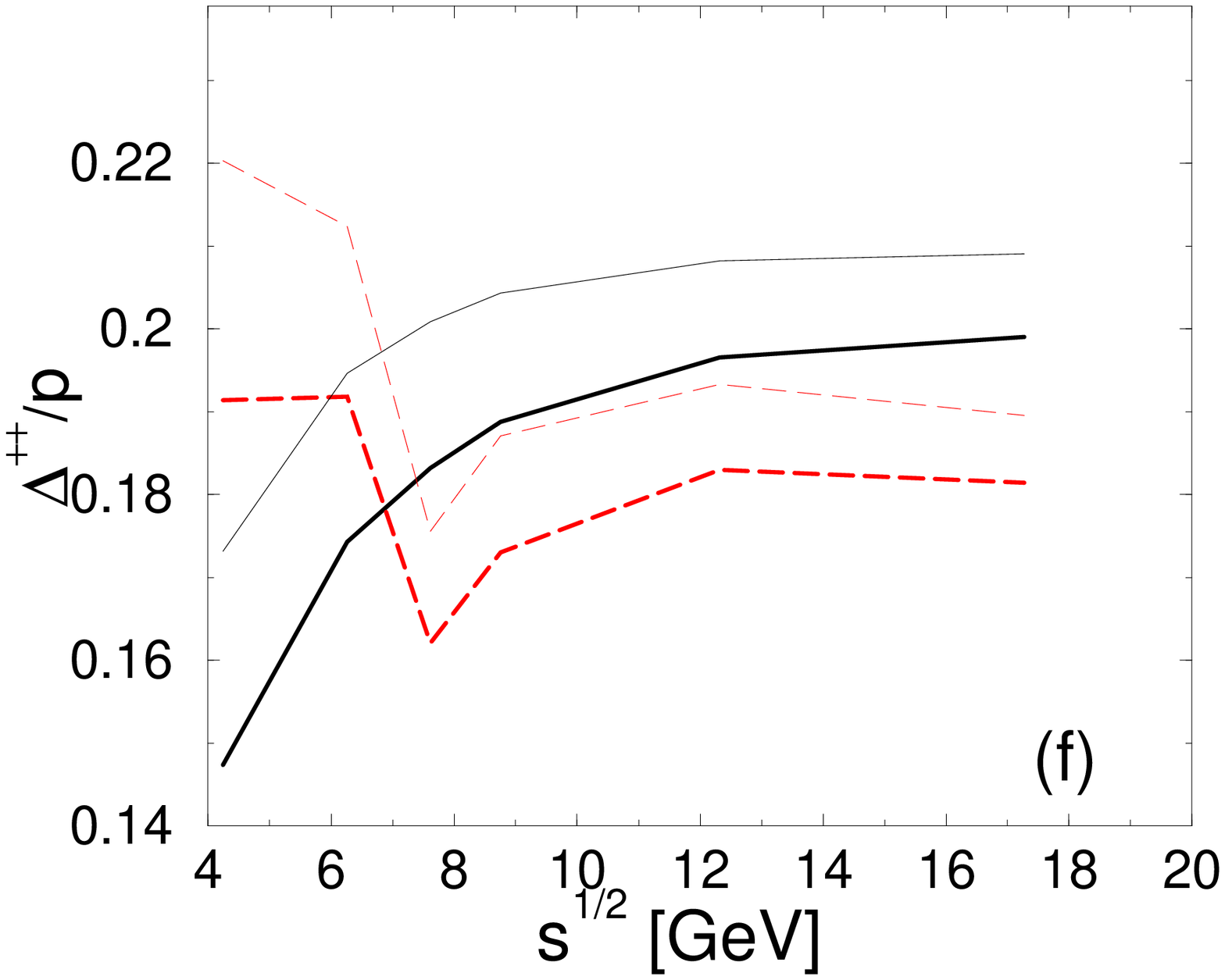}
\caption{(Color online)\label{figres}
Ratio of resonance to the stable particle. Thick lines for particles with strange quark content, 
thin lines for particles with anti-strange quark content,  as a function of energy.
Solid black lines refer to the equilibrium fits ($\gamma_{q,s}=1$), with the parameters for AGS
and SPS energies taken from \cite{equil_energy}.  Dashed red lines refer to
non-equilibrium  fits ($\gamma_{q,s}$ fitted), with the best fit parameters for 
AGS and SPS energies taken from \cite{gammaq_energy}.  }
\end{figure*}

We have evaluated several resonance relative ratios shown in  figure \ref{figres} within the two scenarios, 
using the   statistical hadronization code SHARE \cite{share,sharev2}.
For the non-equilibrium scenario, we have used the parameters given in \cite{gammaq_energy}, 
table I.   For the equilibrium scenario, we used the parametrization given in \cite{equil_energy}  
figures 3 and 4.
In the latter case, the strangeness and isospin chemical potentials $\mu_s$ and $\mu_{I3}$ 
were obtained by requiring that net strangeness be zero, and net charge per baryon in all
particles produced be  the same as in the colliding system.  We have performed spot checks of the
validity of the statistical parameters used and found that under the assumptions made these
are the best parameter sets.

 As  seen    in figure \ref{figres}, the expected trend with $\sqrt{s}$   
is apparent in all  considered resonance ratios, though in
cases where the mass difference is large,
the effect is much more pronounced than in some others. Indeed, 
for many of the ratios we present the experimental error may 
limit the usefulness of our results, however, the opposite energy dependence
may prove to be  helpful in discriminating the behaviour. In principle we have 
presented 12 different
ratios, though in some cases the difference between 
 particle and antiparticle ratio is small. It is smallest in 
cases when the baryochemical potential has neither a direct, nor a large indirect
influence, and the difference in light quark flavor content is the smallest. 

It is interesting to note here that the often  ignored quark flavor effect (isospin effect)
is responsible for most of the difference between particle and antiparticle
ratios. This is at first a counter intuitive result, but it can be understood in 
quantitative simple manner. We recall the relation $u/d\propto \lambda_{I3}\propto \bar d/\bar u$
where $u,d$ refer to the yield of up,  and,  respectively, down valance quarks.
In   cases, such as e.g. $\Sigma^* \rightarrow \Lambda$ the $u,d$ valance  quark  content is different 
for the $R^*$ and $R$ particles, leading to  $\lambda_{I3}$ dependence of the $R^*/R$ ratio.   
High mass cascading resonances, where the {\em strangeness} content  
can be different (e.g.  $\Xi(1690) \rightarrow K \Sigma$), are a further source of difference between particle and antiparticle
resonance ratios, especially so  in the high baryo-chemical potential regime.

Many of the experimental  data points needed in the analysis presented 
in \cite{gammaq_energy} and \cite{equil_energy} are to this day still preliminary.
This means that some of the results we rely on could in the end  be somewhat different.
 However, because of the cancellation (to a good approximation) of the baryo- 
and strangeness chemical  potentials, the qualitative
prediction for the {\em energy dependence} of the resonance yields within 
the two models is robust. Namely, 
within the chemical equilibrium model the 
temperature of chemical freeze-out must steadily increase and so
does the $R^*/R$ ratio. For the chemical non-equilibrium 
model the $R^*/R$ dip primarily relies on the response of $T$   
to the degree of chemical equilibration: prior to chemical equilibrium   
 for the valance quark abundance, at a relatively low reaction energy, the 
freeze-out temperature 
$T$ is relatively high.   At a critical energy, $T$ drops as  the 
hadron yields move to or even exceed light quark chemical equilibrium,
yet reaction energy is still not too large, and thus the baryon density is high
and meson yield low.
As reaction energy increases further, 
$T$ increases  and the $R^*/R$ yield from that point on increases. 
The drop in $R^*/R$ at critical $T$,
would be completely counter-intuitive in an equilibrium picture.
It would hence provide overwhelming evidence that non-equilibrium
effects such as supercooling, where such a drop would be expected, 
are at play.
We further argue that such a drop can not be reproduced by resonance rescattering/regeneration.

 Pseudo-elastic processes such as
$R \pi \rightarrow R^* \rightarrow R\pi $ 
and post-decay $ R^* \rightarrow R\pi $
scattering of decay products in matter 
could potentially considerably alter the observable  final ratio of {\em detectable} $R^*$ to $R$.   
The combined effect of rescattering and regeneration has not been well understood.  
We have argued that the formation  of additional 
{\em detectable} resonances is negligible~\cite{usresonances}, while scattering
of decay products can decrease the visible resonance yields except for 
sudden hadronization case.  Other groups have studied this 
in quantitative manner. 

  Assuming a long lived hadron phase, the
energy dependence of most of the resonance ratios considered here has been calculated in a
hadronic quantum molecular dynamics model. The result   (figure 7 and 8 in~\cite{urqmdreso})
is qualitatively similar to the chemical equilibrium results for resonance ratios, we see a smooth rise with energy.
Thus, in the case of chemical equilibrium, with a considerable separation between chemical and thermal freeze-out
inherent in Ref.~\cite{urqmdreso}
rescattering and regeneration will affect the {\em quantitative} $R^*/R$ ratio, but will not alter
the {\em dependence on heavy ion reaction energy} shown in figure~\ref{figres}.
On the other hand,  chemical
nonequilibrium implies absence of a long lived hadron phase.  Because of this,
the calculation~\cite{urqmdreso} would not be applicable and the resonance abundance
should be in closer quantitative, as well as qualitative agreement, with the predictions of figure~\ref{figres}.
Rescattering and regeneration, therefore, should not alter the predicted pattern of either the
equilibrium model, where all $R^*/R$ should rise with energy, or the non-equilibrium model,
where at the critical energy all $R^*/R$ should experience a dip.

If both of these predictions prove inaccurate, and the $R^*/R$ abundance
turns
out to be resonance specific with no uniform rises and dips as function of
energy, this
would signify that freeze-out is determined by reaction kinetics rather
than
thermodynamic conditions...    In this case, $R^*$ abundance is determined
more by $\Gamma_{R^*} \tau$, where $\tau$ is the lifetime of the fireball,
than by $m_{R^*}/T$.
\section{Conclusions}  
We have argued that a measurement of the energy dependence of ratios
such as $K^*/K$, $\Delta/p$, $\Sigma^*/\Lambda$, $\Lambda(1520)/\Lambda$, $\Xi^*/\Xi$
and other such ratios
might be instrumental in clarifying  the  freeze-out conditions in heavy ion collisions,
especially at low reaction energy.
A resonance abundance monotonically rising with energy from the AGS energy range
would suggest that the best
statistical description of heavy ion data is based on chemical
equilibrium, and that as collision energy increases, freeze-out temperature
rises monotonically.
If, on the other hand, resonance abundance shows a consistent dip, possibly at the energy coinciding with
  the other  non-monotonic  features recently observed in particle yields and ratios \cite{horn},
it would be a strong evidence that what we are seeing is, at and above this dip, a freeze-out
from a supercooled high entropy density phase.
 GT would like to thank 
C. Gale and S.Jeon for helpful discussions.
Work supported in part by grants: from the Natural Sciences and Engineering research 
council of Canada, the Fonds Nature et Technologies
of Quebec,   the U.S. Department of
Energy  DE-FG02-04ER413, and the Tomlinson foundation.

\end{document}